\documentclass[letterpaper, 11pt, notitlepage]{article}
\usepackage[utf8]{inputenc}
\usepackage{amsmath}
\usepackage{amssymb}
\usepackage[usenames,dvipsnames]{xcolor}
\usepackage{graphicx}
\usepackage{float}
\usepackage{pgfplots}
\usepackage{tikz}
\usetikzlibrary{patterns}
\usepackage{titlesec}
\usepackage{mathtools}
\usepackage{capt-of}
\usepackage{array}
\usepackage{booktabs}
\usepackage{rotating}
\usepackage{comment}
\usepackage[top=1in, bottom=1in, left=1in, right=1in]{geometry}

\usepackage{titling}
\usepackage{blindtext}
\newcommand{\pol}{\text{pol}}

\usepackage[style=authoryear, doi=false,isbn=false,url=false,eprint=false,date=year,uniquename=false,backend=biber,natbib]{biblatex}

\renewbibmacro{in:}{}
\usepackage{xpatch}
\xpatchbibmacro{volume+number+eid}{%
  \setunit*{\adddot}%
}{%
}{}{}
\DeclareFieldFormat[article]{number}{\mkbibparens{#1}}

\usepackage{amsthm}
\newtheoremstyle{space}
  {15pt} 
  {15pt} 
  {\normalfont} 
  {} 
  {\bfseries} 
  {.} 
  {.5em} 
  {} 

\theoremstyle{space}
\newtheorem{prop}{Proposition}

\newtheorem*{defn}{Definition}
\newtheorem{axiom}{Axiom}

\def\thm@space@setup{%
  \thm@preskip=5cm
  \thm@postskip=\thm@preskip 
}

\author{\Large Hane Lee \qquad Michael E. Sobel\\  \\ \Large Department of Statistics, Columbia University}
\title{The Wasserstein Bipolarization Index: A New Measure of Public Opinion Polarization, with an Application to Cross-Country Attitudes toward COVID-19 Vaccination Mandates.}

\date{Draft: June 24, 2024}

\begin{document}
\begin{titlingpage}
    \maketitle
    \vspace{30pt}
    \begin{abstract}
    Although the topic of opinion polarization receives much attention from the media, public opinion researchers and political scientists, the phenomenon itself has not been adequately characterized in either the lay or academic literature.  To study opinion polarization among the public, researchers compare the distributions of respondents to survey questions or track the distribution of responses to a question over time using ad-hoc methods and measures such as visual comparisons, variances, and bimodality coefficients. To remedy this situation, we build on the axiomatic approach in the economics literature on income bipolarization, specifying key properties a measure of bipolarization should satisfy: in particular, it should increase as the distribution spreads away from a center toward the poles and/or as clustering below or above this center increases.  We then show that measures of bipolarization used in public opinion research fail to satisfy one or more of these axioms. Next, we propose a $p$-Wasserstein polarization index that satisfies the axioms we set forth. Our index measures the dissimilarity between an observed distribution and a distribution with all the mass clustered on the lower and upper endpoints of the scale. We use our index to examine bipolarization in attitudes toward governmental COVID-19 vaccine mandates across 11 countries, finding the U.S and U.K are most polarized, China, France and India the least polarized, while the others (Brazil, Australia, Colombia, Canada, Italy, Spain) occupy an intermediate position. 
    
    \end{abstract}
\vspace{3.5cm}
\end{titlingpage}

\section{Introduction}
Contemporary media coverage of the American political scene depicts a public sharply divided by different views into opposing camps whose members distrust and dislike those outside their cluster.

Political scientists have offered a more nuanced account. First, they distinguish between political elites and the public, widely agreeing that ``elite polarization'', i.e., ideological homogeneity among party representatives and ideological dissimilarity between representatives of different parties, has steadily increased since the 1970s (\cite{poole2011ideology}).  At the mass level, there is also a consensus that ``partisan sorting'', whereby individuals who self-identify as liberal (conservative) are likely to identify as Democrats (Republicans), has increased (\cite{levendusky2009partisan}; \cite{fiorinaabrams2008}). And there is also agreement that ``affective polarization'', whereby members or supporters of each party view those of the other party negatively, has increased. (\cite{druckman_levy_2022}; \cite{iyengaretal2019}; \cite{SIMAS_CLIFFORD_KIRKLAND_2020}).  

Another strand of literature is concerned with ``ideological polarization'', the dissimilarity between members of the public on policy issues.  Two notions have been espoused (\cite{lelkes2016}).  
``Ideological consistency`` refers to the extent to which persons espouse 
liberal or conservative opinions on multiple issues, whereas   
``ideological divergence'' refers to the distribution of public opinion on given issues, with clustering of respondents at opposing extremes of a univariate scale indicative of polarization on that issue.  


\citet{abramowitz_saunders_2008} equate polarization with ideological consistency (alignment).  Using opinion items from the National Election Study (NES)  they argue that consistency increased over 
the period 1982-2004, as evidenced by both the proportion of respondents giving consistently liberal or conservative responses to the items and the correlation of responses to different items.  
Similarly, the Pew Research Center (2014) reports that the percentage of persons who express ideologically consistent opinions increased from 1994-2014. However, using items from the 1972-2004 NES, \citet{Baldassarri} find little evidence of increased alignment.  Similarly, using items from the General Social Survey (GSS) and the NES, \citet{dimaggio1996} find little evidence of increased ``ideological constraint'' in their analysis spanning 1972-1994.  But a more recent analysis using NES items (\cite{KOZLOWSKI}) points to an increase in correlations among items over from 2004-2016. And in the Canadian context, \citet{merkley2023} finds that the average correlation between eight issue items from the Canadian Election Studies increased moderately from 1993-2019.  

There is widespread agreement that ideological divergence is maximally manifested 
when individuals are concentrated in two clusters at opposite ends of a univariate ordinal or bounded metric scale (\cite{bauer_2019}), suggesting ideological divergence should be measured by comparing the observed distribution of opinion to this exemplar distribution. But researchers have taken a more elliptical approach: using visualization (\cite{Fiorina_Levendusky_2006}; \cite{fiorinaabrams2008}; \cite{koudenberg2021}) and/or ad-hoc measures such as the sample variance and bimodality coefficients (\cite{dimaggio1996}; \cite{strijbis2020cosmopolitan}; \cite{MELKI201436}; \cite{lelkes2016}; \cite{merkley2023}) that appear on the face of it to capture important features of the exemplar, they ask instead whether there is a temporal trend in these measures. 
For example, \citet{fiorinaabrams2008} display figures of distributions intended to help readers visualize polarized vs. non-polarized distributions and they compare frequency distributions of five NES seven point ordinal opinion items in 1984 and 2004: for four of these, the percentage of respondents in the first, seventh and ``central'' categories appears similar at the two time points, indicating to them no change in polarization over this period.  But they do not attempt to say whether or not any of the distributions exhibit polarization, stating that whether or not opinions are polarized at a particular time point ``is generally a matter of judgement'' (p.566). \citet{dimaggio1996} 
argue that increasing the spread and bimodality of a distribution increases ideological divergence. But they do not attempt to formalize and connect these concepts to changes in polarization, simply operationalizing spread by variance, bimodality by kurtosis; 
\citet{mouw_sobel} point out that kurtosis does not measure bimodality.  In their empirical work, DiMaggio et al.\ find no evidence (with the exception of the abortion item) that attitudes have polarized. \citet{hill_tausanovitch_2015} use 67 NES policy items collected in different years between 1956 and 2012, reporting no increase in spread as measured by the variance of estimated ideal points for each respondent. Using NES items from 1972-2012 and Sarle's bimodality coefficent, \citet{lelkes2016} finds no evidence of increasing polarization. Also from NES, \citet{fiorina2017unstable} finds that Americans' self-identified ideological positions have also not polarized between 1972 and 2016.

Economists have also considered polarization.  Here 
the aforementioned disconnect between measurement and conceptualization is bridged through an axiomatic approach in which intuitions about polarization are formalized, and measures conforming to the axioms constructed.  Two strands of literature predominate (\cite{esteban_ray_2012}).  The first (\cite{esteban_ray_1994}; \cite{duclos_esteban_ray}; \cite{esteban_gradin_ray}) views polarization as the grouping of individuals into several (not necessarily two) clusters where members of different groups are possibly antangonistic toward members of other clusters: polarization is then the sum of all these antagonisms.
The second strand, which more closely informs our research, stems from work on the ``hollowing out'' of the ``middle class'' and views polarization, (synonymously bipolarization in this context) as the clustering of  
individuals on opposite sides of the median of a metric scale (e.g, income), with greater distances from the median indicating higher levels of polarization.
Here the notions of spread and bipolarity are axiomatized, leading, in conjunction with functional form and normalization restrictions, to indices that essentially measure the distance of a distribution from its median. 
To the best of our knowledge, statistical properties of these indices have not been explored. See \citet{anderson2004}, who uses stochastic dominance relations to characterize and test for various forms of polarization.  

There is also a related literature on bipolarization for ordinal scales (\cite{ALLISON2004505}; \cite{apouey2007}; \cite{apouey_silber_2013}; \cite{kobus2015}; \cite{sarkar_santra_2020}).  In future work, we shall consider this case: here we simply note that if an investigator knows the relative distance between categories or estimates these using a statistical model that assumes the measurements are cutpoints on an underlying continuum (\cite{mouw_sobel}), the ordinal case can be treated in the same manner as the metric case.  

In section two, axioms that we want an index of polarization to satisfy are set forth; then, a measure of the distance between an observed distribution of opinion and an exemplar consistent with these axioms, the $p$-Wasserstein distance, is proposed.  We also show that commonly used measures such as the variance and Sarle's bimodality coefficent are incompatible with one or more of the axioms.  In section three, the estimated  
$p$-Wasserstein index is used to assess the bipolarization among 11 countries in attitudes toward mandatory COVID-19 vaccination.
The US and UK are the most polarized countries, China, India and France the least, while  Brazil, Australia, Colombia, Canada, Italy and 
Spain are intermediate.  Section four concludes. The appendix describes the statistical theory on which our interval estimates of polarization are based.  An R package (WPI: Wasserstein Polarization Index) was developed to carry out the computations.  

\section{Bipolarization: A New Measure}
We consider responses $X$, with probability measure $\nu_{X}$, to opinion item ${\cal I}$ measured on a metric scale with compact support on $S \subseteq [\ell, L] \subset {\mathbb{R}}^{1}$.
We propose five axioms, two of which formalize the notion that bipolarization increases as the spread and clustering of responses on either side of a central point $c$ increase. Our treatment is closely related to that of \citet{foster_wolfson_2010} and \citet{wang_tsui}, whose indices measure the dissimilarity between $\nu_{X}$ and a measure concentrated at the median $m(\nu_{X})$ of $\nu_{X}$. In contrast, our index, $P(\nu_{X},\xi_{(\nu_{X},c, \gamma)})$,  
measures the distance of $\nu_{X}$, with center $c$ (not necessarily equal to $m(\nu_{X}))$ from a maximally separated measure 
$\xi_{(\nu_{X},c, \gamma)}$ concentrated at the lower and upper poles $\ell$ and $L$.


\begin{defn}
\textbf{The Maximally Separated Measure}. For $X \in [\ell, L]$, with probability measure $\nu_{X}$, the maximally separated measure $\xi_{(\nu_{X},c, \gamma)}$, with center $c \in (\ell, L)$, assigns all probability to the points $\ell$ and $L$ as follows: $\xi_{(\nu_{X},c,\gamma)}(\ell) = \nu_{X}[\ell,c) + \gamma\nu_{X}(c)$, 
$\xi_{(\nu_{X},c,\gamma)}(L) = 1 - \xi_{(v_{X},c,\gamma)}(\ell)$, where $\gamma \in [0,1]$. 
\end{defn}

For the case $v_{X}(c) = 0$, for any $\gamma$, the maximally separated measure results from transferring all the mass of $\nu_X$ to the left of center to $\ell$ and all the mass to the right of center to $L$.  When $v_{X}(c) > 0$, the mass to the left (right) of $c$ is transferred to $\ell$ ($L$), and a proportion $\gamma$, specified by the researcher, of the mass at the center is transferred to $\ell$, the remainder to $L$. If $c$ is chosen to correspond to a quantile 
$q$ of $\nu_{X}$, $\gamma$ is chosen so that the maximally separated measure places mass $q$ on $\ell$ and mass $1-q$ on $L$.

In public opinion research where the variance is used to measure bipolarization, the center is the mean. Substantively motivated choices are also possible.  For items where a point on the scale represents a neutral position, a researcher might want to choose this point as a center.  Public opinion researchers widely agree that the maximum amount of bipolarization occurs when half the responses to an item are clustered at the lower endpoint, the other half clustered at the upper endpoint (\cite{bauer_2019}).  This implies $c$ is the median.  The median also figures prominently in the political science literature, where the ``average citizen'' often refers to the median voter (\cite{Gilens_Page_2014}; \cite{downs}), and the median position historically indicates a ``central tendency among voters'' (\cite{kim_fording}). Similarly, in the economic literature on bipolarization, motivated by concerns over the ``hollowing out of the middle class, the median income represents the center of this class.  
  

It is important to understand the empirical implications of different choices of the center. If the center is chosen to correspond to a quantile $q$, i.e., the center
of $\nu_{X}$ is $c(q)$, $\xi_{(\nu_{X},c(q), \gamma(q))}$ depends only on $\nu_X$ and $q$. But if $c$ is chosen in some other fashion, e.g., as the mean or some point representing a neutral position, 
$\xi_{(\nu_{X},c, \gamma)}$ depends on all three parameters.  



In section 2.1, we propose five axioms we want our index of bipolarization to satisfy. In section 2.2, we show that measures of polarization commonly used in the literature on ideological divergence are incompatible with one or more of these axioms.
In section 2.3, we propose a $p$-Wasserstein measure of bipolarization and prove that it is consistent with the five axioms.


\subsection{Axioms}


Axioms 1 and 2 impose invariance conditions on our index that facilitate comparison across populations and scales.

\begin{axiom}
Let $X$ be a random variable with probability measure $\nu_{X}$ and $Y$ a random variable with probability measure $\nu_{Y}$. Then, if $\nu_{X} = \nu_{Y}$, $P(\nu_{X},\xi_{(\nu_{X},c, \gamma)}) = 
P(\nu_{Y},\xi_{(\nu_{X},c, \gamma)})$ for any choice of $c$ and $\gamma$.
\end{axiom}

Second, the origin $\ell$ and maximum $L$ of items measuring public opinion are arbitrary. For example, the NES feeling thermometer items range from 
``very cold or unfavorable feeling'' ($\ell = 0$) to ``very warm or favorable feeling'' ($L = 100$); political ideology is often assessed on a 10 or 11 point scale running from ``left'' to ``right'' (\cite{bauer_etal_2017}). To facilitate interpretation and enable comparison of responses to one or more items measured on different scales, e.g., the same question asked in different surveys, or to ask whether attitudes toward gun control are more polarized than attitudes toward abortion, we require our index to be translation invariant and homogeneous of degree 1.

\begin{axiom}
Let $X \in [\ell, L]$ denote the response to item ${\cal I}$ in population ${\cal P}$. 
Let $Y = \ell_{Y} + \beta(X - \ell) \in [\ell_{Y}, L_{Y}]$, $\beta = \frac{L_{Y} - \ell_{Y}}{L - \ell}$.  Then
$P(\nu_{Y}, \xi_{(\nu_{Y}, \ell_{Y} + \beta(c - \ell),\gamma)}) = \beta P(\nu_{X}, \xi_{(\nu_{X},c, \gamma)})$.
\end{axiom}

Next, as in the economic literature on bipolarization (\cite{foster_wolfson_2010}, \cite{wang_tsui}, \cite{esteban_ray_2012}) of income, 
we formalize the notions of spread and clustering that are also commonly identified in the literature on ideological divergence as features of polarization.
Motivated by earlier work on the ``hollowing out of the middle class'', \citet{foster_wolfson_2010} characterize increased spread as follows: for responses $X$ and $Y$ measured in the same way, with probability measures $\nu_{X}$ and $\nu_Y$ respectively, and common median $m(\nu_{X}) = m(\nu_{Y}) = m$, $\nu_{Y}$ is more polarized than $\nu_{X}$ if the distribution function $F_{Y}$ 
is stochastically higher than $F_{X}$ to the left of $m$ and $F_{X}$ is stochastically
higher than $F_{Y}$ to the right of $m$.
The intuition
for this definition is that movements from the left (right) of the median further left (right) increase polarization: below we characterize increased spread in terms of such movements. 
\begin{defn}
{\bf Left spread}.  $\nu_{Y}$ is a left spread of $\nu_{X}$ about $c$ if 
$\nu_{Y}(I) = \nu_{X}(I)$ for every sub-interval $I$ of $(c,L]$,
and there exists $x_{1} \in [\ell,c]$ such that $\nu_{Y}(I) \geq \nu_{X}(I)$ for 
every sub-interval $I$ of $[\ell, x_{1})$ and
$\nu_{Y}(I) \leq \nu_{X}(I)$ for every sub-interval $I$ of $(x_{1},c]$.
A left spread about $c$ is said to be restricted if $\nu_{Y}(c) \geq (1 - \gamma)\nu_{X}(c)$.
{\bf Right Spread}.  $\nu_{Y}$ is a right spread of $\nu_{X}$ about $c$ if 
$\nu_{Y}(I) = \nu_{X}(I)$ for every sub-interval $I$ of $[\ell,c)$, 
and there exists $x_{1} \in [c,L]$ such that $\nu_{Y}(I) \leq \nu_{X}(I)$ for every sub-interval $I$ of $[c,x_{1})$ and 
$\nu_{Y}(I) \geq \nu_{X}(I)$ for every sub-interval $I$ of $(x_{1},L]$.
A right spread about $c$ is said to be restricted if $\nu_{Y}(c) \geq \gamma \nu_{X}(c)$.
\end{defn}

The definitions above include the case of a null spread $\nu_{X} = \nu_{Y}$, in which case $x_{1}$ can be any value in $[\ell, c]$ for a left spread about $c$, 
any value in $[c,L]$ for a right spread about $c$. 
For a non-null left spread about $c$, with $x_{1} \in (\ell, c)$, either mass is transferred from $(x_{1},c]$ to $[\ell, x_{1}]$ or from 
$[x_{1},c]$ to $[\ell, x_{1})$;
if $x_{1} = \ell$, $x_{1}$ receives a transfer of mass from $(\ell, c]$, and if $x_{1} = c$, mass is sent from $x_{1}$ toward $\ell$.  
Similarly, for a non-null
right spread about $c$, with $x_{1} \in (c, L)$, either mass is transferred from $[c,x_{1}]$ to $(x_{1},L]$ or from $[c,x_{1})$ to $[x_{1},L]$; 
if $x_{1} = c$, $x_{1}$ sends mass toward $L$, and if $x_{1} = L$, $x_{1}$ receives mass from $[c, L)$.  


\begin{axiom}
If $\nu_{Y}$ is a restricted left (restricted right) spread of $\nu_{X}$ about $c$, 
$P(\nu_{Y},\xi_{(\nu_{X},c, \gamma)}) \leq P(\nu_{X},\xi_{(\nu_{X},c, \gamma)})$.
\end{axiom}
The restriction $\nu_{Y}(c) \geq (1 - \gamma) \nu_{X}(c)$ on a left spread about $c$ ensures that at most $\gamma \nu_{X}(c)$ probability mass is moved from $c$ toward $\ell$; otherwise, it would be necessary to transfer some mass in $[\ell,c)$ to $c$ in order to attain the distribution corresponding to the maximally separated measure. Similarly, the restriction $\nu_{Y}(c) \geq \gamma \nu_{X}(c)$ on a right spread about $c$ ensures that at most $(1 - \gamma) \nu_{X}(c)$ probability mass is moved from $c$ toward $L$.

A second feature of bipolarization is clustering: moving masses on the left (right) of center closer together should increase polarization.  But this may also increase spread.  Consider the case of a discrete distribution $\nu_{X}$, where $\nu_{X}(x_{1}) = p_{1} > 0$, $x_{1} < x_{2} \leq c$, $\nu_{X}(x_{2}) = p_{2} > 0$.   
A new measure $\nu_{Y}$ is created by moving 1) 
mass $p_{2}^{*} < p_{2}$ from $x_{2}$ to $x_{2} - \delta$ and 2) mass $p_{1}^{*} < p_{1}$ from $x_{1}$ to $x_{1} + \eta$, where $x_{1} < x_{1} + \eta \leq 
x_{2} - \delta < x_{2}$.  Clearly, move 1 increases spread and move 2 decreases spread, and $p^{*}_{1}$, $p^{*}_{2}$, $\delta$, and $\eta$ may be chosen so that the net spread either increases or decreases. To capture changes in clustering independently of spread, these quantities should be chosen so that the net change is 0, i.e., the mean value $E(Y)$ associated with $\nu_{Y}$ equals the mean value $E(X)$ associated with $\nu_{X}$. 
To formalize this requirement, we first define a ``mean preserving merge'', then adapt this to the case of clustering on either side of $c$.    Our definition is adapted from the definition of a mean preserving spread
in \citeauthor{Machina1997IncreasingRS} (1997; 105-106), who generalized the definition of a mean preserving spread in \citet{ROTHSCHILD1970225}. 

\begin{defn}
\textbf{Mean preserving merge}. Let $\nu_{X}$ and $\nu_{Y}$ be probability distributions on $[\ell, L]$ with common mean $E(X) = E(Y)$.  
We say $\nu_{Y}$ is obtained from $\nu_{X}$ by a mean preserving merge (equivalently, $\nu_{X}$ is obtained from $\nu_{Y}$ by a mean preserving spread) if there exist values $x_{1} \leq x_{2}$ such that : a) $\nu_{Y}(I) \leq \nu_{X}(I)$ for every sub-interval $I$ of $[\ell, x_{1})$, 
b) $\nu_{Y}(I) \geq \nu_{X}(I)$ for every sub-interval $I$ of $(x_{1},x_{2})$, c) $\nu_{Y}(I) \leq \nu_{X}(I)$ for every sub-interval $I$ of $(x_{2}, L]$.
\end{defn}

\indent  The case $x_{1} = x_{2}$ occurs for a null merge $\nu_{X} = \nu_{Y}$ or if mass is sent from $[\ell, x_{1})$ and $(x_{1}, L]$ to $x_{1}$. 
The point $x_{1}$ can transfer probability to or from $\nu_{X}(x_{1},x_{2})$ and the point $x_{2}$ can transfer probability to or from $\nu_{X}(x_{2},L]$.
The Pigou-Dalton transfer, used by Wang and Tsui (2002) to characterize clustering, whereby an individual with income $x_{1} > x_{0}$  transfers amount $\lambda < x_{1} -x_{0}$ to the individual with income $x_{0}$, is a special case of the mean preserving merge.  We now adapt the definition above to characterize clustering, in which masses ``below'' or ``above'' $c$ are moved closer together. 

\begin{defn}
\textbf{Left Merge}. Let $\nu_{Y}(I) = \nu_{X}(I)$ for every sub-interval $I$ of $(c,L]$.  If $\nu_{Y}$ is a mean preserving merge of $\nu_{X}$ with $x_{2} \leq c$, we say $\nu_{Y}$ is a left mean preserving merge about $c$.  We say a left mean preserving merge about $c$ is restricted if $\nu_{Y}(c) \geq (1 - \gamma) \nu_{X}(c)$. \textbf{Right Merge}. Let $\nu_{Y}(I) = \nu_{X}(I)$ for every sub-interval $I$ of $[\ell,c)$.  If $\nu_{Y}$ is a mean preserving merge of $\nu_{X}$ with $x_{1} \geq c$, we say $\nu_{Y}$ is a right mean preserving merge about $c$. We say a right mean preserving merge about $c$ is restricted if $\nu_{Y}(c) \geq \gamma \nu_{X}(c)$.  
\end{defn}

\begin{axiom}
If $\nu_{Y}$ is obtained from $\nu_{X}$ by a restricted left mean preserving merge about $c$ or a restricted right mean preserving merge about $c$,
$P(\nu_{Y},\xi_{(\nu_{X},c, \gamma)}) \leq P(\nu_{X},\xi_{(\nu_{X},c, \gamma)})$.
\end{axiom}

Our goal is to develop an index that measures the dissimilarity between $\nu_{X}$ and its maximally separated counterpart 
$\xi_{(\nu_{X},c, \gamma)}$ in which the mass to the ``left'' of center is transferred to $l$ and mass to the ``right'' to $L$.  We require our measure to be a distance: 


\begin{axiom}
For any two measures $\nu_{X}$ and $\nu_{Y}$, $P(\nu_{X}, \nu_{Y})$ is a distance.
\end{axiom}



\subsection{Polarization Indices in Public Opinion Research: Variance and Bimodality}

Public opinion researchers typically use either variance or Sarle's bimodality coefficient to measure polarization. Neither of these satisfy all the axioms we have set forth. First, recall that a mean preserving merge is the reverse of a mean preserving spread, and it is well known 
that a mean preserving spread increases variance.  Thus, a mean preserving merge reduces variance, so  
using the variance to measure polarization would suggest, contrary to Axiom A4, that increased clustering is associated with decreased bipolarization.  Next, we examine Sarle's finite sample bimodality coefficient (Lelkes 2016):
\begin{align*}
    b=\frac{g^2+1}{k+\frac{3(n-1)^2}{(n-2)(n-3)}},
\end{align*}
where $g=\frac{m_3}{s^3}$ is the sample skewness,  $k=\frac{m_4}{s^4}-3$ is the sample excess kurtosis, $m_3$ and $m_4$ are the sample third and fourth central moments, and $s$ is the sample standard deviation. Clearly, Axiom A1 is violated, as two samples with different sizes and the same values of $k$ and $g$ yield different values of $b$. While this is inconsequential for ``large'' $n$, consider next two samples of size $n$ with observed probability measures $\nu_X= 0.5\cdot\delta_{0.4} + 0.5\cdot\delta_{0.6}$ and $\nu_Y = 0.5\cdot\delta_{0.1} + 0.5\cdot\delta_{0.9}$ on $[0,1]$, where $\delta_x$ denotes a Dirac measure with mass 1 at $x$. Clearly, $\nu_Y$ exhibits increased spread compared to $\nu_{X}$, but $g$ and $k$ are equal, and $b$ will suggest, at least in large samples, that $\nu_{X}$ and $\nu_{Y}$ are equally polarized, violating Axiom A3. \\


\subsection{$p$-Wasserstein measure of polarization}


We propose to measure the distance between $\nu_{X}$ and $\xi_{(\nu_{X},c, \gamma)}$ using the $p$-Wasserstein distance, and we show that axioms A1-A5 are consistent with 
this choice.  Our definition is adapted from \citet{villani2009optimal}: 

\begin{defn}
\textbf{$p$-Wasserstein distance}.
Let $(\mathcal{X}, d)$ be a Polish metric space, and let $p \in [1,\infty)$. For any two probability measures $\nu_X, \nu_Y$ on $\mathcal{X}$, the Wasserstein distance of order $p$ is defined by
\begin{equation}
    W_p(\nu_X, \nu_Y)= \displaystyle{ \left(\inf_{\pi \in \Pi(\nu_X, \nu_Y)} \int_{\mathcal{X}}d(x,y)^p d\pi(x,y) \right)^{\frac{1}{p}}}, 
\label{eq:wasserstein}
\end{equation}
where $\Pi(\nu_X, \nu_Y)$ is the set of all joint probability measures on $\mathcal{X} \times \mathcal{X}$ with marginals $\nu_X$ and $\nu_Y$, respectively. 
\end{defn}

\noindent For a given metric $d$, (\ref{eq:wasserstein}) is the minimum cost of ``movement'' needed to transform the distribution $\nu_{X}$ into $\nu_{Y}$ or $\nu_{Y}$ into $\nu_{X}$ (\cite{solomon2018optimal}).  Here we take $d(x,y) = |y - x|$. 

Our choice of the Wasserstein distance is motivated by it capacity to capture ``key geometric properties of the underlying ground space'' that other statistical distances do not (\cite{peyre}). Consider Figure 1, with $c$ equal to the median $m(\nu_{X})$, 
and denote the most polarized distribution, with masses .5 at $\ell$ and $L$, $\xi_\pol$. Let $\nu_Y$ be obtained from $\nu_{X}$ by shifting the mass of $\nu_{X}$ to the right of $c$ $k$ units to the right (a right spread). Intuitively, $\nu_{Y}$ exhibits greater polarization than  $\nu_{X}$: further, to satisfy axiom A3 our measure must satisfy $D(\nu_{Y},\xi_\pol)<D(\nu_{X},\xi_\pol)$, as will be the case using the $p$-Wasserstein distance. However, other commonly used distances, such as the total variation distance between probability measures $\nu_{X}$ and $\nu_{Y}$ ($D_\text{TV}(\nu_{X}, \nu_{Y}) \coloneqq \sup_{A\subset \mathcal{X}} |\nu_{X}(A)-\nu_{Y}(A)|$), and the Hellinger distance ($D_\text{H}(\nu_{X}, \nu_{Y})\coloneqq\frac{1}{2}\int_\mathcal{X} ( \sqrt{\nu_{X}(dx)}-\sqrt{\nu_{Y}(dx)}) ^2$) will not discriminate between these cases because neither $\nu_{X}$ nor $\nu_{Y}$ share any points of common support with $\xi_\pol$: nor do these distances satisfy Axioms A2-A4.  And the Kullback-Leibler (KL) divergence  ($D_\text{KL}(\nu_{X}, \nu_{Y})\coloneqq\int_\mathcal{X} \log\left(\frac{\nu_{X}(dx)}{\nu_{Y}(dx)}\right) \nu_{X}(dx)$), another popular measure of the dissimilarity between distributions, albeit not a distance, is not applicable, as $\nu_{X}$ ($\nu_{Y}$) and $\xi_\pol$ do not share a common support. \\

\pgfmathdeclarefunction{gauss}{2}{%
  \pgfmathparse{1/(#2*sqrt(2*pi))*exp(-((x-#1)^2)/(1*#2))}%
}

\begin{center}
    \resizebox{7cm}{!}{%
    
\begin{tikzpicture}
\begin{axis}[
  no markers, domain=0:10, samples=100,
  xmin=0, xmax=10, axis lines*=left, axis y line=none, 
  every axis y label/.style={at=(current axis.above origin),anchor=south},
  every axis x label/.style={at=(current axis.right of origin),anchor=west},
  height=5cm, width=12cm,
  xtick=\empty, ytick=\empty,
  enlargelimits=false, clip=false, axis on top,
  grid = major,
  ]
  \addplot [fill=CornflowerBlue!15,  draw=none, domain=3:7] {gauss(5,0.8)} \closedcycle;
  \addplot [domain=3:7, thick, CornflowerBlue!85!black] {gauss(5,0.8)};
  \addplot [domain=3:5, thick, Periwinkle] {gauss(5,0.8)};
  \addplot [pattern=crosshatch dots, pattern color=Periwinkle, draw=none, domain=6.8:9.3] {gauss(6.8,0.8)} \closedcycle;
  \addplot [domain=6.8:9.3, thick, Periwinkle] {gauss(6.8,0.8)};
  \addplot [pattern=crosshatch dots, pattern color=Periwinkle, draw=none, domain=3:5] {gauss(5,0.8)} \closedcycle;

  \draw[very thin, Periwinkle] (50,0) -- (50,500);
  \draw[very thin, Periwinkle] (68,0) -- (68,500);
  \draw [](axis cs:0.2,-0.05) node [fill=white] {\large $\ell$};
  \draw [](axis cs:9.8,-0.05) node [fill=white] {\large $L$};
  \draw [](axis cs:5,-0.057) node [fill=white] {\large $c$};
  \draw [](axis cs:6.8,-0.05) node [fill=white] {\large $c+k$};
  \draw [](axis cs:5.9,0.43) node [fill=white, text=CornflowerBlue!85!black] {\Large $\nu_X$};
  \draw [](axis cs:8,0.3) node [fill=white, text=Periwinkle] {\Large $\nu_Y$};
\end{axis}
\end{tikzpicture}
}

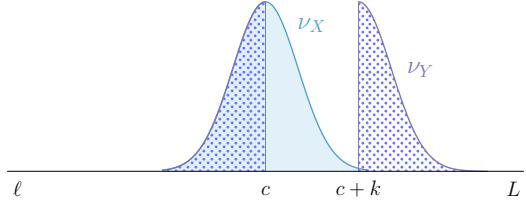
\captionof{figure}{$\nu_Y$ is more polarized than $\nu_X$.}
\end{center}


Calculating the Wasserstein distance requires finding the optimal coupling between two measures that minimizes the cost of transport. Suppose for the moment $\nu_X(c)=0$.  It seems intuitively obvious that the optimal coupling $\pi_\text{opt}$ between $\nu_X$ and $\xi_{(\nu_{X},c,0)}$ with $d(x,y) = |y - x|$ moves mass to the left of $c$  further left to $\ell$ and mass to the right of $c$ further right to $L$. Any other coupling $\pi \not= \pi_{\text{opt}}$ will map at least some mass $\epsilon>0$  from the left of $c$, say from $x \in [\ell,c)$, to $L$ and the same amount of mass from the right of $c$, say from $x'\in(c,1]$, 
to $\ell$. Then, the distance traveled by the $\epsilon$ masses increases from $(L-x')+(x-\ell)$ under $\pi_{\text{opt}}$  to $(L-x')+2(x'-x)+(x-\ell)$ under $\pi$, and the cost increases accordingly. We establish this claim more generally for the $p$-Wasserstein distance in Proposition 1:

\begin{prop}
\label{prop:optcoupling}
Let $\nu_X$, with center $c$, be a probability measure defined on $\mathcal{X}=[\ell,L]$, and let $\xi_{(\nu_{X},c, \gamma)}$ denote the corresponding maximum separation measure. Then, the optimal $p$-Wasserstein coupling $\pi_\text{opt}$ maps mass $\nu_{X}[\ell,c) + \gamma\nu_{X}(c)$ to $\ell$, mass $1 - \{ \nu_{X}[\ell,c) + \gamma\nu_{X}(c)\}$ to $L$.
\end{prop}


\noindent Proof. Let $\pi_{\text{opt}}$ denote the optimal coupling between $\nu_X$ and $\xi_{(\nu_{X}, c, \gamma)}$. Any other coupling $\pi \not= \pi_{\text{opt}}$ will move at least equal amounts of mass $\epsilon$ from $A \subset [\ell,c]$ to $L$ and from $B \subset [c, L]$ to $\ell$. Let $\nu_A$ denote the Borel measure constructed from the $\epsilon$ mass on $A$ and $\nu_B$ the measure constructed from the $\epsilon$ mass on $B$. In other words,  $\nu_A$, $\nu_B$ are measures such that $\nu_A(A)=\nu_B(B)=\epsilon$, $\nu_A(A')\leq\nu_X(A')~\forall A' \in A-\{c\}$, $\nu_A(c)\leq \gamma\nu_X(c)$; $\nu_B(B')\leq\nu_X(B')~\forall B' \in B-\{c\}$, and $\nu_B(c)\leq (1-\gamma) \nu_X(c)$.
\begin{align}
    D^p_{\pi}(\nu_X,& \xi_{(\nu_{X},c, \gamma)}) - D^p_{\pi_\text{opt}}(\nu_X, \xi_{(\nu_{X},c, \gamma)})  \nonumber \\
    &=\int_A (L-x)^p d\nu_A + \int_B (x-
    \ell)^p d\nu_B - \int_A (x-\ell)^p d\nu_A - \int_B (L-x)^p d\nu_B \label{align:diff}
\end{align}
Now,  for the mass on $A$,
\begin{equation}
    \int_A (L-x)^p d\nu_A - \int_A (x-\ell)^p d\nu_A \geq \epsilon((L-c)^p  - (c-\ell)^p),
\label{eq:A} 
\end{equation}
with equality if and only if mass $\epsilon$ is moved from $c$ to $L$ under $\pi$. Equivalently, for the mass on $B$,
\begin{equation}    
    \int_B (x-\ell)^p d\nu_B- \int_B (L-x)^p d\nu_B \geq \epsilon((c-\ell)^p - (L-c)^p),
\label{eq:B}  
\end{equation}
with equality if and only if mass $\epsilon$ is moved from $c$ to $\ell$ under $\pi$.

Since moving equal amounts of mass $\epsilon$ from $c$ to $L$ in (\ref{eq:A}) and from $c$ to $\ell$ in (\ref{eq:B}) reproduces $\pi_{\text{opt}}$,
for any other coupling $\pi$, either the inequality in (\ref{eq:A}) or (\ref{eq:B}) is strict, hence (\ref{align:diff}) $ > 0$. This inequality also indicates that the optimal coupling is unique. $\square$\\

Substantively, the optimal coupling $\pi_\text{opt}$, by moving people left (right) of $c$ to the left (right) extreme, captures the idea of public opinion change by ``movements'' of individuals across a ``spatial'' spectrum of political ideology (\cite{Zaller_1992}; \cite{downs}), inducing a state of polarization as people move ``away from the center toward the extremes'' (Fiorina and Abrams 2006), with ``liberals (conservatives) gravitating more reliably to the liberal (conservative) position'' (Zaller 1992 p.102).

Next, we show that the Wasserstein distance satisfies axioms A1-A5.

\begin{prop}
    \label{prop:axiom}
    $W_p(\nu_X, \xi_{(\nu_{X},c, \gamma)})$ satisfies axioms A1-A5.
\end{prop}

\noindent Proof. Axioms A1 and A5 are satisfied as (\ref{eq:wasserstein}) is a distance between probability distributions. Axiom A2 follows from the definition of the Wasserstein distance with $d(x,y)=|y-x|$:  when $X^*=a+ bX$ and $Y^*=a+bY$ for $a, b \in \mathbb{R}$, 
$W_p(\nu_{X^*}, \nu_{Y^*}) =  |b|W_p(\nu_{X}, \nu_{Y})$.  It remains only to show that Axioms A3 and A4 are satisfied. Because Axiom A2 is satisfied, we proceed, without loss of generality, taking $[\ell,L]=[0,1]$.



To prove Axiom A3 is satisfied, consider first a restricted left spread about $c$:
\begin{align}
    W^p_p(\nu_{X} &, \xi_{(\nu_{X},c, \gamma)}) - W^p_p(\nu_{Y}, \xi_{(\nu_{X},c, \gamma)}) \nonumber \\
    &=  \int_{[0, x_{1})}x^p d\nu_{X} -\int_{[0, x_{1})}x^p d\nu_{Y} + x_{1}^{p}(v_{X}(x_{1}) - v_{Y}(x_{1})) + 
    \int_{(x_{1},c]}x^p d\nu_{X} -\int_{(x_{1},c]}x^p d\nu_{Y} \nonumber \\ 
    & \geq x_{1}^{p}[(\nu_{X}[0,x_{1})) - \nu_{Y}[0,x_{1})) + (\nu_{X}(x_{1}) - \nu_{Y}(x_{1})) + (\nu_{X}(x_{1},c] - \nu_{Y}(x_{1},c])] \nonumber \\
    & = 0,
\label{align:A3}
\end{align}
The case of a restricted right spread about $c$ is proved in an analogous fashion.

\indent   
That Axiom A4 is satisfied follows from Theorem 3 in \citet{Machina1997IncreasingRS}, which states that the conditions a) $\nu_{Z}$ is obtained from  $\nu_{X}$ by a sequence of mean preserving spreads and b) $\int_{[0, 1]} u(x)d\nu_{Z} \leq \int_{[0,1]}u(x)d\nu_{X}$ for every concave function $u(\cdot)$ are equivalent.
For a restricted left mean preserving merge about $c$
\begin{equation}
    W^p_p(\nu_{X},  \xi_{(\nu_{X},c, \gamma)}) - W^p_p(\nu_{Y}, \xi_{(\nu_{X},c, \gamma)}) 
    = \int_{[0,1]}x^p d\nu_{X} -\int_{[0,1]}x^p d\nu_{Y} = 
    \int_{[0,c]}x^p d\nu_{X} -\int_{[0,c]}x^p d\nu_{Y}. 
\label{eq:A4}
\end{equation}
For $p = 1$, the definition of a mean preserving merge implies (\ref{eq:A4})$ = 0$. For $p > 1$
the result follows from the definition of a restricted left mean preserving merge about $c$ and the convexity of $x^{p}$ on $[0,1]$. 
For a restricted right mean preserving merge about $c$, the result follows similarly from the definition and the convexity 
of $(1 - x)^{p}$ on $[0,1]$. $\square$ \\

\section{A Cross-Country Comparison of Attitudes toward Government Mandates on Vaccination for COVID-19}

\begin{sidewaysfigure}[hbtp]
    \centering
    \includegraphics[width=\textwidth]{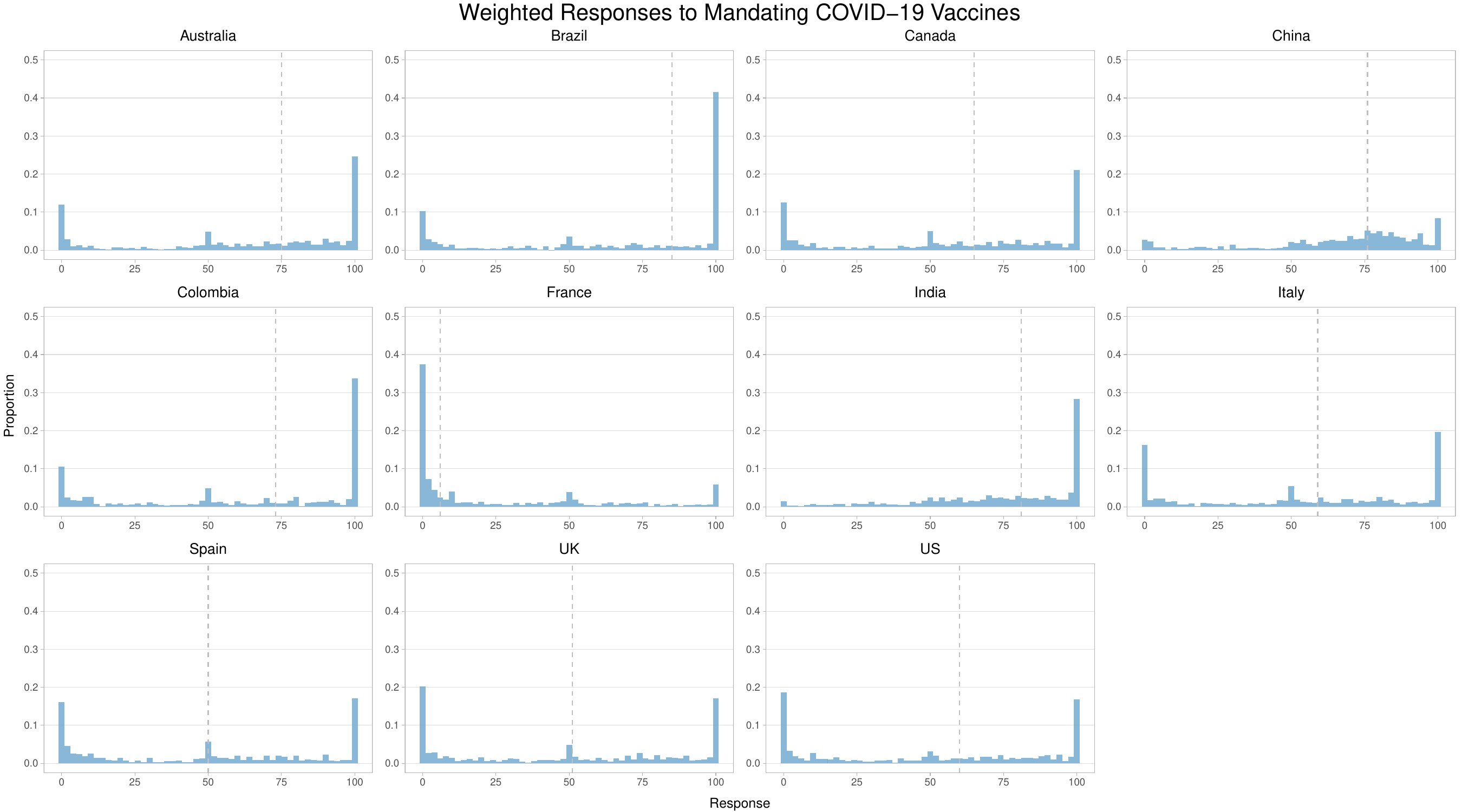}
    \caption{Each histogram features a set of responses from one country to the question, ``The government should make the COVID-19 vaccine mandatory for everybody,'' measured on a 0 (Disagree) to 100 (Agree) VAS scale. Dashed lines indicate the median.}
    \label{fig:hist}
\end{sidewaysfigure}

The COVID-19 pandemic prompted unprecedented governmental actions such as lockdowns, international travel restrictions, and vaccination mandates that stirred much public debate wihin and across countries. Debates about the safety and efficacy of vaccines, the prioritization of groups 
during rollout, and, especially, vaccination mandates, were often contentious, and the substance of these disagreements varied across countries.  
For instance, Democrats and Republicans clashed over vaccination in the US, lead by conflicting messages from Biden and Trump (\cite{BOLSEN202281}). And in Brazil, where many citizens strongly supported vaccination, followers of (at the time) President Bolsonaro, a COVID-19 denier, displayed significantly lower support (\cite{GRAMACHO20212608}).  On the other hand, Chinese opinions reflected ``stabilized public sentiment'', possibly due to timely government responses (\cite{han2020using}).

To assess the variation in public opinion on vaccine allocation policies, \citet{duch_candour} conducted the Oxford COVID-19 Vaccine Preference and Opinion Survey (CANDOUR), an online survey of 15,536 adults 18 years of age or more, administered between November 2020 and January 2021 in 13 countries, with 1000 to 1500 respondents from each country. In Chile and Uganda, excluded from our analysis, respondents were recruited through Facebook.  In the remaining countries (Australia, Brazil, Canada, China, Colombia, France, India, Italy, Spain, the UK, and the US) sampling was conducted by the survey research firm Respondi using respondents from their online panels. Except for Canada and Spain, where demographic and spatial characteristics of sample respondents were close to population totals, post-stratification weights calculated through raking were used to adjust for remaining imbalances between sample and population marginals on region, age, sex and education (\cite{duch_candour}; see online appendix). 

We apply our index to a question from the survey that asked respondents how much they agree with the following statement: ``The government should make the COVID-19 vaccine mandatory for everybody.'' Responses were recorded as integers on a 0 (disagree) to 100 (agree) visual analogue scale (VAS). Weighted histograms of responses from each country are presented in Figure \ref{fig:hist}.

\subsection{Rescaling $W_p(\nu_{X},  \xi_{(\nu_{X},c, \gamma)})$ to $[0,1]$: The index $I_p(\nu_{X},  \xi_{(\nu_{X},c, \gamma)})$.}

To facilitate interpretability, we rescale the $p$-Wasserstein distance $W_p(\nu_{X}, \xi_{(\nu_{X},c, \gamma)})$ to lie in $[0,1]$ so that the lowest (highest) level of polarization corresponds to a 0 (1) value of the index. To do so we derive the upper and lower bounds of $W^p_p(\nu_{X}, \xi_{(\nu_{X},c, \gamma)}).$  The lower bound for $W^p_p(\nu_{X}, \xi_{(\nu_{X},c, \gamma)})$ is obviously $0$, which occurs if and only if 
$\nu_{X} = \xi_{(\nu_{X},c, \gamma)}$, corresponding to an index value
$I_p(\nu_{X},  \xi_{(\nu_{X},c, \gamma)}) = 1$. As the upper bound depends on $[\ell, L]$ (see Axiom A2) we assume $X$ has already been transformed to lie in $[0,1]$.

\begin{prop}
    \label{prop:upperbound}
     Let $Q(q) = \inf \{x: \nu_X[0,x]\geq q\}, q \in [0,1]$, denote the quantile function associated with $\nu_{X}$, 
    and let 
     $\xi_{(\nu_{X},c, \gamma)} \equiv \xi_q=q\cdot \delta_0 + (1-q) \cdot \delta_1$, where $\delta_x$ is a Dirac measure with mass 1 on $x$. Then,
    \begin{center}
        $W_p(\nu_X, \xi_q) \leq W_p(\delta_{Q(q)}, \xi_q) \leq \max(q^{\frac{1}{p}}, (1-q)^{\frac{1}{p}})$.
    \end{center}
    
\end{prop}

\noindent Proof:
\begin{align*}
    W_p^p(\nu_X, \xi_q) &= \int_{[0, Q(q))} x^p d\nu_X + \gamma \nu_X(Q)\cdot Q(q)^p + \int_{(Q(q), 1]} (1-x)^p d\nu_X + (1-\gamma) \nu_X(Q) \cdot [1-Q(q)]^p\\
    &\leq \int_{[0, Q(q))} Q(q)^p d\nu_X + \gamma \nu_X(Q)\cdot Q(q)^p + \int_{(Q(q), 1]} [1-Q(q)]^p d\nu_X + (1-\gamma) \nu_X(Q) \cdot [1-Q(q)]^p \\
    &= q \cdot [Q(q)]^p + (1-q) \cdot [1-Q(q)]^p = W_p^p(\delta_{Q(q)}, \xi_q) \\
    & \leq \max(q, 1-q) = \max[W_p^p(\delta_1, \xi_q), W_p^p(\delta_0, \xi_q)]. \quad \square
\end{align*}

Thus, we define the index $I_p(\nu_{X},  \xi_{(\nu_{X},c, \gamma)}) = 1 - [\max(q^{\frac{1}{p}}, (1-q)^{\frac{1}{p}})]^{-1}  
W_p(\nu_{X},  \xi_{(\nu_{X},c, \gamma)})$. If $q < .5$ ($q > .5$) the index takes the minimum value of $0$ when 
$W_{p}^{p}(\nu_{X},\xi_{q}) =  W_{p}^{p}(\delta_{0},\xi_{q}) = 1 - q$ ($W_{p}^{p}(\nu_{X},\xi_{q}) =  W_{p}^{p}(\delta_{1},\xi_{q}) = q$), i.e., there is a consensus of opinion at 0 (1). 


Now, for any value of $q \in [0,1]$, let 
$Q_{p}^{*}(q) = \arg \min_{\{0 \leq Q(q) \leq 1\}}W_{p}(\delta_{Q(q)},\xi_{q})$ denote the value of $Q(q) \in [0,1]$ that maximizes 
$I_{p}(\delta_{Q(q)},\xi_{q})$. It follows from above that $Q_{1}^{*}(q) = 1$ if $q <.5$, 0 if $q > .5$ and $\{0,1\}$ if $q = .5$.    For $p > 1$
$Q_{p}^{*}(q) = \bigl(\frac{1-q}{q}\bigr)^{1/(p-1)}/1 +\bigl(\frac{1-q}{q}\bigr)^{1/(p-1)}$, $p > 1$. Further, for $p \geq 1$,
$W_p(\delta_{x}, \xi_q)$ increases monotonically as $|x - Q_{p}^{*}(q)|$ increases.  
Thus, in contrast to several indices of income bi-polarization that take the median $m(\nu_{X})$ as the center and result in a value of $0$ when the distribution is concentrated on $m(\nu_{X})$, no matter where this median is located (Foster and Wolfson 2010; Wang and Tsui 2002), $I_p(\delta_{Q(q)}, \xi_q)$ is sensitive to the location where mass is concentrated,  
with location at one or both ends of the scale (depending on the choice of $q$) evidencing less polarization than locations at intermediate points on the scale. These properties of our index allow us to capture important features of opinion polarization that indices which do not depend on the location of points of concentration miss, for example the observation that opinions located at either end of the opinion spectrum are more stable and difficult to change than opinions that are less extreme (\cite{druckman_leeper_2012}, \cite{Zaller_1992}).

\subsection{Results}

As previously noted, public opinion researchers typically regard the most polarized state as that in which half of the responses are located at $\ell$, the other half at $L$. This implies a center $c$ equal to the median $m(\nu_{X})$, where any choice of $c$ with $P(X\leq m(\nu_X))\geq 0.5$ and $P(X\geq m(\nu_X))\leq 0.5$ will result in the same optimal coupling, with half of the responses at $\ell$ and half at $L$.  After rescaling the responses to lie in the unit interval $[0,1]$, we define $\xi_\pol=0.5\cdot\delta_0+0.5\cdot\delta_1$ and estimate the index 
$I_p(\nu_X, \xi_\pol)=1-2^{1/p}\cdot W_p(\nu_X, \xi_\pol)$.
In our illustration, we treat the responses as measurements on an interval scale: if a researcher does not believe a difference of $x_{1} - x_{2} = \Delta$ units on the scale represents the same amount of opinion as a difference on the scale of $x_{3} - x_{4} = \Delta$ units, he may wish to apply a monotone  
non-linear transformation to the original values so that the transformed values more nearly meet this assumption.  Second, the asymptotic confidence intervals we report are obtained under the assumption of simple random sampling (\cite{sommerfeld_phd}); to compute these, the R package WPI (Wasserstein Polarization Index) 
was developed. 

\begin{figure}[h]
    \centering
    \includegraphics[width=0.8\textwidth]{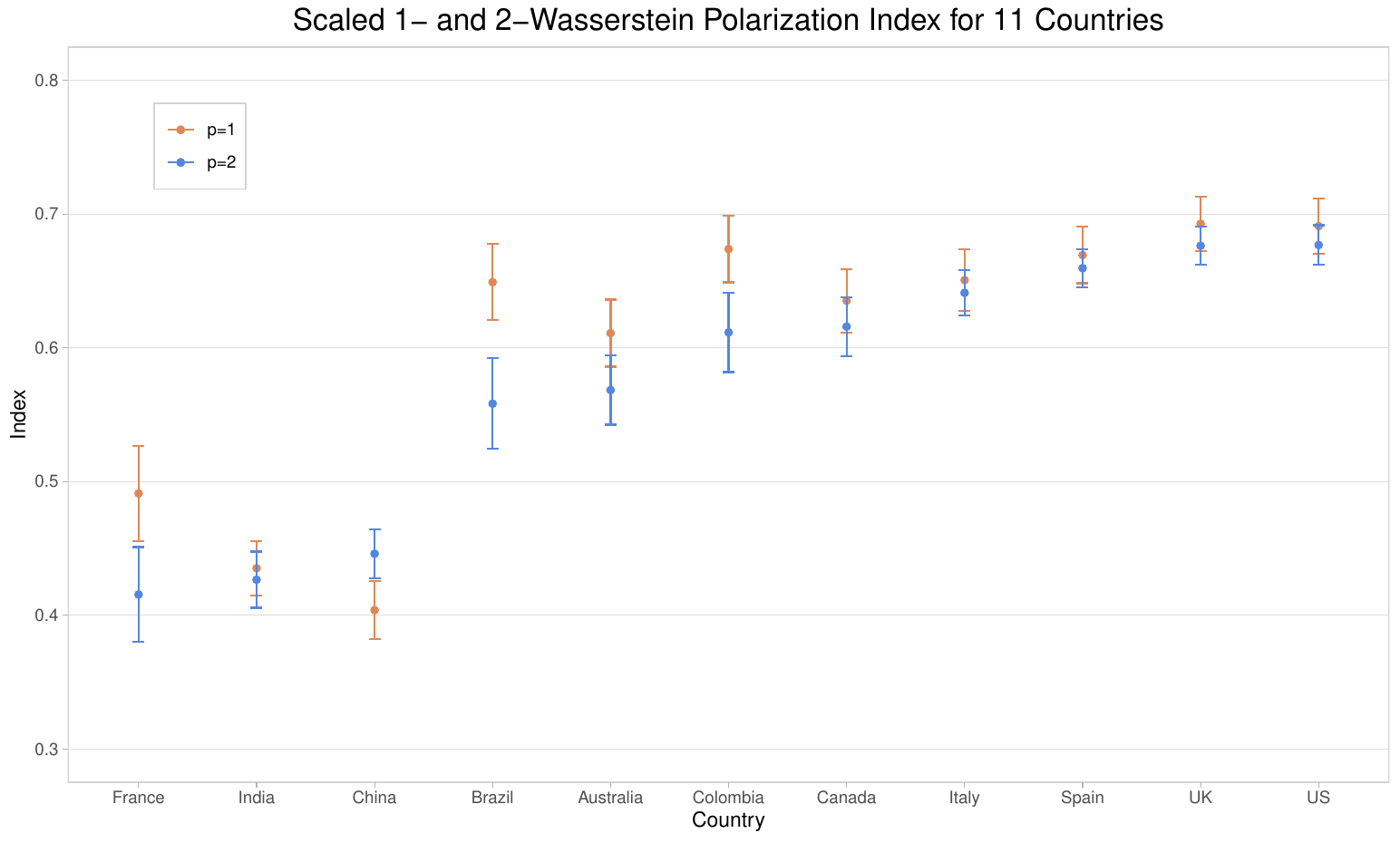}
    \caption{$p=1$ and $p=2$ Wasserstein polarization index point estimates with bars indicating 95\% confidence intervals. Countries are ordered by increasing polarization for $p=2$.}
    \label{fig:CI}
\end{figure}

In their analysis of opinions regarding mandatory vaccination, 
Duch et al.\ do not define or measure polarization.  Using visual inspection of the histograms in Figure \ref{fig:hist} they classify some of the 11 countries into three groups:
``highly polarized'' (US and UK), where the majority of respondents are ``either strongly opposed or strongly supportive'', 
``somewhat polarized'' (Australia, Brazil, and Colombia), where there is ``little middle ground'', 
and ``broad consensus'' (China, India, and France).  Despite this lack of rigor, their grouping is roughly congruent with the ordering of our estimates
$\widehat{I_{p}}(\nu_{X},\xi_{pol})$ for $p =1$ and $p =2$. However, our index is not only based on an explicit notion of polarization, but our estimates  allow for more refined, quantitative comparisons that take uncertainty into consideration. For instance, within the ``somewhat polarized'' group, only Australia is significantly less polarized than the ``highly polarized'' group for $p=1$. More generally, the eight countries that do not belong to the ``broad consensus'' group cannot be readily clustered into groups without arbitrary divisions. We also provide point estimates and confidence intervals for Canada, Italy, and Spain, three countries that Duch et al.\ do not attempt to classify. 

Though our country ordering is essentially the same for both $p =1$ and $p=2$, there are several differences.
While China, India and France are the least polarized countries, France appears less polarized than China for $p=2$, but more polarized for $p=1$.  All else equal, if France and China had equal amounts of clustering around 0 and 0.75, then the distribution clustered around 0.75 would display more polarization for $p=2$ but not $p=1$. Furthermore, France does not exhibit much clustering except near 0,  whereas China exhibits stronger clustering on both sides of the median. This also increases polarization for $p=2$, but not $p=1$. A similar reversal occurs between Brazil and Australia, but here, for both $p = 1$ and for $p = 2$ the 95\% confidence intervals for the two countries overlap.

\section{Discussion}
To study polarization empirically,  
public opinion researchers use ad hoc measures that fail to capture key components of the intuitions they have about the characteristics of polarization. Building on the economic literature on bipolarization,  we set out key properties an index of bipolarization should have and propose a $p$-Wasserstein index that incorporates these properties. 
To illustrate our index, we re-analyze an item from a study by Duch et al.\ (2021) that asked respondents in a number of countries whether they thought that vaccination against COVID-19 should be mandatory in their country.  Among the countries studied, China, India and France are the least polarized, while the UK and US are the most polarized.

Our results are based on cross country comparisons of interval estimates. Similarly, comparing the distribution of items tracked over time, albeit typically informally, a number of studies have asked whether or polarization has increased in the recent past  (\cite{dimaggio1996}; \cite{fiorinaabrams2008}; \cite{mouw_sobel}; \cite{lelkes2016}); our index could be used to address this question as well. It would be useful to develop frameworks for modeling trends in our index and for testing hypotheses about group differences in index values.  

Second, many studies are based on scales with seven or fewer response categories. Most public opinion researchers have treated responses from such scales as interval data.  An exception is Mouw and Sobel (2001), who treated a four point scale measuring attitudes toward abortion as ordinal and asked whether the distribution of an underlying continuous latent attitude toward abortion had changed over time.  Analogues to the indices of bipolarization developed by economists for continuous data have been developed by economists working with ordinal self-assessed health data (Allison and Foster 2004; Apouey 2007; Apouey and Silber 2013). By using a different cost metric to measure the difficulty of moving between different categories, it is straightforward to extend our index to this case; more generally we shall consider the ordinal case in greater detail in future work.


Third, in line with the literature on ideological divergence, we studied polarization with respect to individual opinion items.  Researchers studying ideological consistency (alignment, constraint) have also argued that polarization is reflected in the relationship between different items. However, common measures of consistency are ad hoc. For example, for each respondent, \citet{abramowitz_saunders_2008} use 7 NES items, subtracting the number of liberal responses from the number of conservative responses, then using the absolute value of the difference to construct a consistency scale ranging from 0 to 7; the conclusion that ideological consistency has increased is then obtained by comparing the proportion of respondents with scale values of 4 or greater over time.    
Using a similar scale constructed from 10 items, the  \citet{pew2014} concludes that ideological consistency increased between 1994 and 2014.  Others have argued that increased consistency is (also) evidenced by increased correlations among items over time  (\cite{abramowitz_saunders_2008}; \cite{Baldassarri};
\cite{dimaggio1996}; \cite{KOZLOWSKI}; \cite{merkley2023}).  To operationalize this notion, \citet{merkley2023} codes policy items as liberal or conservative, then tracks the value of the average correlation among the items over time, whereas \cite{dimaggio1996} operationalize 
``ideological constraint'' using Cronbach's alpha.  
As an alternative to these approaches, one might attempt to develop a more principled ``multivariate'' polarization index, a possibility we wish to explore in future work.

To the best of our knowledge, ours is the first paper to use the theory of optimal transport to develop a sociopolitical index.  Whereas our index is defined as a comparison of a distribution $\nu_{X}$ 
to a prescribed distribution with mass concentrated on the minimum and maximum scale values, the Wasserstein distance is frequently used to compare arbitrary distributions $\nu_{X}$ and $\nu_{Y}$. In social research, where comparisons between two groups are ubiquitous (e.g., incomes of men and women, educational attainments of blacks and whites) the disparity between groups is often operationalized as a difference in means or medians of an outcome of interest; the Wasserstein distance, with its capacity to capture the relative geometry between distributions, is able to detect dissimilarities between
distributions that are missed by these cruder summaries.  Thus, we believe the Wasserstein distance has a potentially important role to play in social research for conceptualizing, measuring and testing between group similarities and dissimilarities.  As an immediate application, we intend to use the Wasserstein distance to study  affective polarization, a topic of great interest in the lay and academic literature, often conceptualized by political scientists and public opinion researchers as the ``spatial'' distance between partisans of different parties and measured as a mean difference between the respondent's affect on a visual analog scale (feeling thermometer) toward the party to which he belongs and a party to which he does not, e.g., Democrats vs. Republicans  (\cite{iyengaretal2019}; \cite{mccarty2019polarization}; \cite{enders2021issues}). 

\printbibliography

\appendix
\section*{Appendix: Computation and Inference}


The $p$-Wasserstein distance between discrete measures $\nu_X$ and $\nu_Y$ on a finite support $\mathcal{X}$ can be expressed as:
\begin{align*}
    [W_p(\nu_X, \nu_Y)]^p = \inf_{\pi \in \Pi(\nu_X,\nu_Y)} \sum_{x,y\in \mathcal{X}}d^p(x,y)\pi(x,y),
\end{align*}
where $\Pi(\nu_X, \nu_Y)$ is the set of all joint probability measures with marginals $\nu_X$ and $\nu_Y$. Let $\{x_i\}_{i=1}^M$ denote the support of $\nu_X$ and $\{y_j\}_{j=1}^N$ the support of $\nu_Y$. Then the two measures can be expressed as ``superpositions''  $\nu_X=\sum_{i=1}^M p_{i}\delta_{x_i}$ and $\nu_Y=\sum_{j=1}^N q_j \delta_{y_j}$ (\cite{solomon2018optimal}), and the $p$-Wasserstein distance can be computed by the following linear program.
\begin{align*}
    \min_{\boldsymbol{\pi}} \quad &\sum_{i,j}d^p(x_i,y_j)\pi_{ij}\\
    \text{s.t.} \quad & \sum_i \pi_{ij} = q_j, \sum_j \pi_{ij} = p_i, \pi_{ij}\geq0.
\end{align*}
The dual of this linear program is given by:
\begin{align*}
	\max_{(\mathbf{v},\mathbf{w}) \in \mathbb{R}^M \times \mathbb{R}^N} \mkern3mu & \sum_{i=1}^M v_ip_i + \sum_{j=1}^N w_j q_j\\
	\text{s.t.}\quad \mkern17mu & v_i + w_j \leq d^p(x_i, y_j).
\end{align*}
with dual variables $\mathbf{v}, \mathbf{w}$.



Using the dual, \citet{sommerfeld_phd} derive the asymptotic distribution of the empirical Wasserstein distance between $\hat{\nu}_n$, the empirical measure of $\nu$ with $n$ observations, and a known measure $\xi$ (Theorem 8(a)), from which approximate confidence intervals directly follow: \\ \\
Assume $\hat{\nu}_n, \xi$ on $\mathcal{X}$ satisfy the non-degeneracy condition: $\sum_{x \in A} \hat{\nu}_n(x) \not = \sum_{y \in B} \xi(y)$ for all proper subsets $A, B$ of $\mathcal{X}$. Let  $\mathbf{v}^*, \mathbf{w}^*$ denote a solution to the dual problem of $W_p(\hat{\nu}_n, \xi)$. Then, an approximate $(1-\alpha) \times 100\%$ confidence interval for $W_p(\hat{\nu}_n, \xi)$ is given by
\begin{equation*}
    \left[W_p(\hat{\nu}_n, \xi) - \frac{z_{\alpha/2}}{\sqrt{n}p}W_p^{1-p}(\hat{\nu}_n, \xi)\sigma(\hat{\nu}_n, \xi), W_p(\hat{\nu}_n, \xi) + \frac{z_{\alpha/2}}{\sqrt{n}p}W_p^{1-p}(\hat{\nu}_n, \xi)\sigma(\hat{\nu}_n, \xi) \right],
\end{equation*}

\noindent where $\quad \sigma ^2 (\hat{\nu}_n, \xi) = \sum_{x \in \mathcal{X}} \mathbf{v}^*(x)^2 \hat{\nu}_n(x) -   \left( \sum_{x \in \mathcal{X}} \mathbf{v}^*(x) \hat{\nu}_n(x) \right)^2 $and $z_{\alpha/2}$ is the $(1 - \alpha/2)$ quantile of the standard normal distribution. Therefore, for our rescaled index  $I_p(\hat{\nu}_n, \xi)=1-\max(q^{1/p},(1-q)^{1/p})^{-1}\cdot W_p(\hat{\nu}_n, \xi)$, the confidence interval is

\begin{equation*}
\begin{split}
    \Biggl[ 1-\max(q^{1/p},(1-q)^{1/p})\Bigl\{ W_p(\hat{\nu}_n, \xi) &+ \frac{z_{\alpha/2}}{\sqrt{n}p}W_p^{1-p}(\hat{\nu}_n, \xi)\sigma(\hat{\nu}_n, \xi) \Bigr\}, \\ 
    &1- \max(q^{1/p},(1-q)^{1/p})\Bigl\{W_p(\hat{\nu}_n, \xi) -  \frac{z_{\alpha/2}}{\sqrt{n}p}W_p^{1-p}(\hat{\nu}_n, \xi)\sigma(\hat{\nu}_n, \xi) \Bigr\}\Biggr].
\end{split}
\end{equation*}


Note that the degeneracy condition follows from linear programming assumptions. The code used to compute the confidence intervals in the empirical application is given in the R package ``WPI'': Wasserstein Polarization Index.  

Theorem 8(a) covers the case of the distance between $\nu_{x}$ and the known measure $\xi$, as in our application.  Theorem 8(b) (\cite{sommerfeld_phd}) covers the two sample case. 



\end{document}